\documentclass[twocolumn,showpacs,preprintnumbers,amsmath,amssymb]{revtex4}
\usepackage{graphicx}% Include figure files
\usepackage{dcolumn}% Align table columns on decimal point
\usepackage{bm}% bold math

\def\be{\begin{equation}}
\def\ee{\end{equation}}
\def\bear{\begin{eqnarray}}
\def\eear{\end{eqnarray}}
\def\bLambda{\mbox {\boldmath ${\Lambda}$}}

\def\bGamma{\mbox {\boldmath ${\Gamma}$}}

\def\bPi{\mbox {\boldmath ${\Pi}$}}
\def\bbeta{\mbox {\boldmath ${\beta}$}}

\include{boldsymb}

\begin{document}

\preprint{APS/123-QED}

\title{Yukawa Fluids
 in the Mean Scaling Approximation:\\ III New Scales }
\author{L. Blum}
 \email{lesblum@yahoo.com}
\affiliation{
Department of Physics, P.O. Box 23343, University of Puerto Rico, Rio, Piedras, PR 00931-3343
}

\author{J.A. Hernando}
 \affiliation{Department of Physics, Comision Nacional de Energia,
Atomica, Av. del Libertador 8250, 1429 Buenos Aires, Argentina.}

\date{\today}
\begin{abstract}

In recent work a general solution of the Ornstein Zernike equation for a general Yukawa closure for a single component fluid was found. Because of the complexity of the equations a simplifying assumption was made, namely that the main scaling matrix $\bGamma$ had to be diagonal. While in principle this is mathematically correct, it is not physical because it will violate symmetry conditions when  different Yukawas are assigned to different components. In this work we show that by using the symmetry conditions the off diagonal elements of $\bGamma$ can be computed explicitly for the case of two Yukawas, and that although the solution is different than in the diagonal case, the excess entropy is formally the same as in the diagonal case. Analytical expressions for the Laplace transforms of the pair distribution functions are derived.
 
\end{abstract}

\maketitle

\section{\label{sec:level1}Introduction}

 There are many problems of practical and academic interest
that can be formulated as closures of some kind of either scalar or matrix Ornstein-Zernike (OZ) equation. These closures can always be expressed by a sum of exponentials, which do form a complete basis set if we allow for complex numbers \cite{blmub,blmhe3}.

While the initial motivation was to study simple approximations like the
Mean Spherical (MSA) or Generalized Mean Spherical Approximation (GMSA), the
availability of closed form solutions for the general closure of the hard
core OZ equation makes it possible to write down analytical solutions for
any given approximation that can be formulated by writing the direct
correlation function $c(r)$ outside the hard core as

\be
c(r)=\sum_{n=1}^{M}K^{(n)}e^{-z_{n}(r-\sigma )}/r 
=\sum_{n=1}^{M}{\cal K}^{(n)}e^{-z_{n}r}/r\label
{eq:yk1} 
\ee
 In this equation $K^{(n)}$ is the interaction/closure constant 
used in the general solution  first found by Blum and Hoye (which we 
will call BH78) \cite
{bh78}, while ${\cal K}^{(n)}$ is the definition used in the later 
general solution by Blum, Vericat and Herrera (BVH92 in what follows) 
\cite{bvh92}. In this work we will use the more common notation of BVH92. 
The case of factored interactions discussed by Blum, \cite{b80} was simplified by
Ginoza \cite{gin1,gin2,gin5,gin6} who found that as in the case of
electrolytes \cite{b3} the solution of the one exponent case could be
expressed in terms of a single scaling parameter $\Gamma $. In the
factorizable case we have 
\be
{ K}^{(n)}=K^{(n)}\delta
_{i}^{(n)}\delta ^{(n)}
\qquad {\cal K}^{(n)}=K^{(n)}d_{i}^{(n)}d^{(n)}\label{eq:yk2} 
\ee
where we have defined 
\be
\delta _{i}^{(n)}=d_{i}^{(n)}e^{-z_{n}\sigma _{i}/2}\label{eq:yk2a} 
\ee

The general solution of this problem was formulated in by Blum, Vericat and
Herrera \cite{bvh92} in terms of a scaling matrix $
\bGamma $. The full solution was given recently by Blum et al. \cite
{blmub,blmhe1,blmhe2}. For only one component the matrix $
\bGamma $ was assumed  to be diagonal diagonal and explicit expressions for the closure relations
for any arbitrary number of Yukawa exponents $M$ were obtained. The
solution is  then remarkably simple in the MSA since then explicit formulas for
the thermodynamic properties are obtained.\\

The diagonal assumption is however not correct for mixtures, even if they are of the same  hard core diameter. In this work we use the symmetry relations to  calculate explicitly the off diagonal terms of $\bGamma$ in the 1 component, 2 yukawa case.
\section{Summary of Previous Work}
\label{prework}

We study the Ornstein-Zernike (OZ) equation 
\be
h_{ij}(12)=c_{ij}(12)+\sum_{k}\int d3h_{ik}(13)\rho _{k}c_{kj}(32)\label%
{eq:oz} 
\ee
where $h_{ij}(12)$ is the molecular total correlation function and $%
c_{ij}(12)$ is the molecular direct correlation function, $\rho _{i}$ is the
number density of the molecules i, and $i=1,2$ is the position $\vec{r}_{i}$
, $r_{12}=|\vec{r}_{1}-\vec{r}_{2}|$ and $\sigma _{ij}$ is the distance of
closest approach of two particles $i,j$. The direct correlation function is

\be
c_{ij}(r)=\sum_{n=1}^{M}K_{ij}^{(n)}e^{-z_{n}(r-\sigma _{ij})}/r,\qquad
r>\sigma _{ij}\label{eq:msac} 
\ee

\noindent and the pair correlation function is 
\be
h_{ij}(r)=g_{ij}(r)-1=-1,\qquad r\leq \sigma _{ij}\label{eq:msac1} 
\ee

We use the Baxter-Wertheim (BW) factorization of the OZ
equation

\be
\left[ {\bf I}+{\bf {\rho }\tilde{H}(k)}\right] \left[ {\bf I}-{\bf \rho }%
\tilde{{\bf C}}(k)\right] ={\bf I}\label{eq:oz2} 
\ee
where $I$ is the identity matrix, and we have used the notation 
\be
\tilde{{\bf H}}(k)=2\int_{0}^{\infty }dr\cos (kr){\bf J}(r)\label{eq:oz2a} 
\ee
\be
\tilde{{\bf C}}(k)=2\int_{0}^{\infty }dr\cos (kr){\bf S}(r)\label{eq:oz2b} 
\ee

The matrices $J$ and $S$ have matrix elements

\be
J_{ij}(r)=2\pi \int_{r}^{\infty }dssh_{ij}(s)\label{eq:oz3a} 
\ee
\be
S_{ij}(r)=2\pi \int_{r}^{\infty }dssc_{ij}(s)\label{eq:oz3b} 
\ee

\be
\left[ {\bf I}-{\bf \rho }\tilde{{\bf C}}(k)\right] =\left[ {\bf I}-{\bf 
\rho }\tilde{{\bf Q}}(k)\right] \left[ {\bf I}-{\bf \rho }\tilde{{\bf Q}}
^{T}(-k)\right] \label{eq:factor} 
\ee
where $\tilde{{\bf Q}}^{T}(-k)$ is the complex conjugate and transpose of $
\tilde{{\bf Q}}(k)$. The first matrix is non--singular in the upper half
complex $k$-plane, while the second is non--singular in the lower half
complex $k$-plane.

It can be shown that the factored correlation functions must be of the form 
\be
\tilde{{\bf Q}}(k)={\bf I}-{\bf \rho }\int_{\lambda _{ji}}^{\infty }dre^{ikr}
\tilde{{\bf Q}}(r)\label{eq:q(k)q(r)} 
\ee

\noindent where we used the following definition 
\be
\lambda _{ji}={\frac{1}{2}}(\sigma -\sigma _{i})\label{eq:q7} 
\ee

{ 
\be
{\bf S}(r)={\bf Q}(r)-\int dr_{1}{\bf Q}(r_{1})\rho {\bf Q^{T}}(r_{1}-r)\label{eq:sq(r)} 
\ee
}

Similarly, from Eq. (\ref{eq:factor}) and Eq. (\ref{eq:oz2}) we get, using
the analytical properties of $Q$ and Cauchy's theorem 
\be
{\bf J}(r)={\bf Q}(r)+\int dr_{1}{\bf J}(r-r_{1}){\bf \rho }{\bf Q}(r_{1})%
\label{eq:jq(r)} 
\ee

The general solution is discussed in \cite{b80,gin1}, and yields

\be
q_{ij}(r)=q_{ij}^{0}(r)+\sum_{n=1}^{M}D_{ij}^{(n)}e^{-z_{n}r}\qquad \lambda
_{ji}<r\label{eq:p1} 
\ee
\begin{eqnarray}
q_{ij}^{0}(r)&=&(1/2)A[(r-\sigma /2)^{2}-(\sigma _{i}/2)^{2}]+ \nonumber  \\
&  &\beta[(r-\sigma /2) -(\sigma _{i}/2)] + \nonumber \\
&  &\sum_{n=1}^{M}C_{ij}^{(n)}e^{-z_{n}\sigma /2}[e^{-z_{n}(r-\sigma
/2)}-e^{-z_{n}\sigma _{i}/2}];  \nonumber \\
&  &\qquad \lambda _{ji}<r<\sigma _{ij}  \label{eq:p2} 
\end{eqnarray}
From here

\bear
{X}_{i}^{(n)}-\sigma _{i}\phi _{0}(z_{n}\sigma
_{i}){\Pi}_{i}^{(n)}\nonumber
 =\\\delta _{i}^{(n)}-\frac{1 }{2}\sigma
_{i} \phi_0(z_n \sigma_i)\sum_\ell \rho_\ell \beta^0_\ell X^{(n)}_\ell-\sigma _{i}^3 z_n^2\psi_1(z_n \sigma_i)\Delta ^{(n)}\nonumber \\
\label{mew2}
\eear
or

\be
\sum_{\ell }{\rho_\ell}\left\{-
 \hat{\cal J}_{j \ell  }^{(n)}\Pi _{\ell }^{(n)}+
\hat{\cal I}_{j \ell }^{(n)}
X_{\ell}^{(n)}\right\}
=\delta ^{(n)}
\label{s11fzz} 
\ee
\subsection{ The Laplace Transforms}

From Eq. \ref{eq:jq(r)} we obtain
the Laplace transform of the pair correlation function 
\be 
2 \pi\sum_{\ell} {\tilde g}_{i{\ell}}(s) [\delta_{\ell j}- \rho_{\ell} {\tilde q}_{{\ell}j}(i s)]=
{\tilde q}^{0^{'}}_{ij}(i s)
\label{eq:p18}
\ee
where  
\[ 
\tilde{q}^{0^{'}}_{ij}(i s)=\int_{\sigma_{ij}}^{\infty} dr 
e^{-s r}[q^0_{ij}(r)]'
\]
\bear
=\left[\left(1+\frac{s \sigma_i}{2}\right) A_j+
s \beta_j\right]\frac{e^{-s \sigma_{ij}}}{s^2}-  \nonumber \\
\sum_m \frac{z_m}{s+z_m} e^{-(s+z_m) \sigma_{ij}}C^{(m)}_{ij}
\label{eq:p20}
\eear
The Laplace transform of Eqs.(\ref{eq:p1}) and  (\ref{eq:p2}) yields

\begin{eqnarray} 
e^{s \lambda_{ji}}\tilde{q}_{ij}(i s)=
\sigma_i^3\psi_1( s\sigma_i)A_j +\sigma_i^2\phi_1( s\sigma_i)\beta_j + \\ \nonumber
\sum_m \frac{1}{s+z_m}[(C_{ij}^{(m)}+ D_{ij}^{(m)})e^{-z_m \lambda_{ji}}
  \\ \nonumber
-C_{ij}^{(m)}e^{-z_m \sigma_{ji}} z_m \sigma_i\phi_0( s\sigma_i)C_{ij}^{(m)} e^{-z_m \sigma_{ji}}]
\label{eq:p20f}
\end{eqnarray}
This result will be used below.

Another important relation deduced from Eq.[\ref{eq:p18}]by setting
\[
s=z_n
\]
 is 
\be
-\Pi^{(n)}_j=\sum_m \tilde{M}_{nm}a^{(n)}_j 
\label{aclos}
\ee
where
\be
\tilde{M}_{nm}=\frac{1}{z_n+z_m}  
 \sum_{\ell} \rho_{\ell}\left[X_{\ell}^{(n)} ( z_m X_{\ell }^{(m)}-\Pi_{\ell }^{(m)})+  X_{\ell}^{(m)} \Pi_{\ell}^{(n)}\right]
\label{defM}
\ee

\section{ The  General Closure}
\label{closure} 
The closure relation (BVH92 \cite{bvh92} ) is, for only one component
\bear
2\pi K\delta ^{(n)}/z_{n}+a^{(n)}{\cal I}^{(n)} 
-\sum_{m}\frac{1}{z_{n}+z_{m}}\rho a^{(n)}a^{(m)} \nonumber
\\ \left[ {\cal J}^{(n)}[\Pi ^{(m)}-z_{m}X^{(m)}]-{\cal I}^{(n)}X^{(m)}\right] 
=0
\label{eqs14f} 
\eear

For the one component case this simplifies to
 \be
  \sum_m \left\{ 2 \pi K \delta_{nm}+
  z_n \Lambda^{(nm)}+\frac{z_n}{z_n+z_m}[ \rho a^{(n)} a^{(m)}]
  \right\} X^{(m)} =0
  \label{cloff2a}
 \ee
 
This can also be written as

 \bear 
   2 \pi K 
 \rho[ X^{(n)}]^2 +
  z_n \rho  a^{(n)} X^{(n)} \nonumber \\
  +\sum_m\frac{z_n}{z_n+z_m}[ \rho a^{(n)} a^{(m)}]  \left[ \rho  X^{(m)} X^{(n)} \right]=0
  \label{beta3a}
 \eear
which is the desired expression. This equation
  simplifies to \cite{blmhe3}
 \be
  2 \pi \rho K_n  \left[ X^{(n)}\right]^2+ z_n \beta^{(n)}\left[1+
  \sum_m \frac{1}{z_n+z_m}   \beta^{(m)} \right]
 =0
 \ee
where $ \beta^{(n)} $ is 
\be
\beta^{(n)}=\rho  X^{(n)} a^{(n)}
\label{beta1c}
\ee

\section{SYMMETRY}

In this section we will summarize and extend our previous analysis of the most general scaling relation \cite{bvh92} for the multiyukawa closure of the Ornstein Zernike equation. We have
\be
\Pi_{i}^{(n)}=- \sum_m \Gamma_{nm} X_{i}^{(m)}
\label{pi2x}
\ee
where $\Gamma_{mn}$ is the $M \times M$ matrix of scaling parameters. This matrix is not uniquely defined by the MSA closure relations and must be supplemented by $M(M-1)$ equations obtained from symmetry requirements for the correlations. 
From  the symmetry of the direct correlation function  at 
the origin, Eq. (\ref{eq:sq(r)}) 
\be
{q}_{ij}(\lambda_{ji})={q}_{ji}(\lambda_{ij})
\ee
we write
\be
a_{i}^{(n)}= \sum_m \Lambda_{nm} X_{i}^{(m)}
\label{eq:s18f}
\ee
where, as was shown in reference \cite{bvh92}, $\bLambda$ must be a symmetric matrix.\\

From the symmetry of the contact pair correlation function Eq. (\ref{eq:jq(r)})
we get
\be
 \{g_{ij}(\sigma_{ij})=g_{ji}(\sigma_{ij})\}\Longrightarrow  \{q_{ij}(\sigma_{ij})'=q_{ji}(\sigma_{ij})'\}
\ee
which are
\be
\sum_n(\Pi_{i}^{(n)}-z_n X_{i}^{(n)})a^{(n)}= \sum_n(\Pi^{(n)}-z_n X^{(n)})a^{(n)}_{i}
\label{eq:s20f}
\ee
from which we get the scaling relation
\be
\Pi_{i}^{(n)}-z_n X_{i}^{(n)}= \sum_ m\Upsilon_{nm}a^{(m)}_{i}
\label{eq:s21f}
\ee
and a new set of $M(M-1)/2$ symmetry relations
\be
 \Upsilon_{mn} =\Upsilon_{nm}
\label{eq:s22f}
\ee

Furthermore, using the scaling relations  we get
\be
{\bf{\tilde M}} \cdot \bf{ \Lambda}={\bf\Gamma}
\label{eq:s23f}
\ee
where the matrix ${\bf{\tilde M}}$ (see Eq.[\ref{defM}])has elements
\be
[{\bf{\tilde M}}]_{nm} = \frac{1}{s_{nm}}
\sum_{\ell} \rho_{\ell} \left[X_{\ell}^{(n)}\{ z_m  X_{\ell }^{(m)}\Pi_{\ell }^{(m)}\}+ X_{\ell }^{(m)}\Pi_{\ell}^{(n)} \right]
\label{eq:s24f}
\ee
Solving these equations yields the relations
\be
{\bf{\tilde M}}^{-1}\cdot {\bf\Gamma}=\bf{\Lambda}
\label{eq:s25f}
\ee
and
\be
-\left(\bf{I}+ \bf{z} \cdot{\bf\Gamma}^{-1}\right)\cdot 
{\bf{\tilde M}} =\bf{\Upsilon}
\label{eq:s26f}
\ee
Both $\bf{\Upsilon}$ and $\bf{\Lambda}$ must be symmetric matrices.
We have therefore a total of $M(M-1)$ symmetry relations, which together with the M closure equations  give the required
equations for the $M^2$ elements of the matrix ${\bf\Gamma}$.\\
The symmetry requirements are more explicitly

\be {\bf\Gamma}\cdot {\bf{\tilde M}}^{T}={\bf{\tilde M}}\cdot{\bf\Gamma}^{T} \qquad {\bf{\tilde M}}^{T}\cdot\left[{\bf\Gamma}^{T}\right]^{-1}={\bf\Gamma}^{-1}\cdot{\bf{\tilde M}} \qquad  {SI}
\label{SI}
\ee
and
\be
\left(\bf{I}+ \bf{z} \cdot {\bf\Gamma}^{-1}\right)\cdot {\bf{\tilde M}} ={\bf{\tilde M}}^{T}\cdot \left(\bf{I}+  [{\bf\Gamma}^{-1}]^{T} \cdot \bf{z}\right)\qquad {SII}
\label{SII}
\ee

 the matrix ${\bf{\tilde M}}$ as
\be
{\bf{\tilde M}}=\frac{1}{2}{\bf{\tilde D}}+\frac{1}{2}{\bf{\tilde M}^A}
\label{mmat}
\ee
and
\begin{widetext}
\bear
[{\bf{\tilde M}^A}]_{nm}=\frac{-1}{s_{nm}}
\sum_{\ell} \rho_{\ell} \left[X_{\ell}^{(n)} ( z_m X_{\ell }^{(m)}-2\Pi_{\ell }^{(m)})- (z_n X_{\ell}^{(n)} -2\Pi_{\ell}^{(n)})X_{\ell }^{(m)}\right]
\nonumber \\
\qquad=
\frac{-1}{z_n+z_m}
\sum_{p} \rho \left[X^{(n)} X^{(p)}(z_m \delta_{p m}+2\Gamma_{ m p }) -(z_n \delta_{p n}+2\Gamma_{n p })X^{(m)} X^{(p)}\right]
\eear
\end{widetext}
\be
[{\bf{\tilde M}^A}]_{nm}=-X^{(n)} X^{(m)}\left[\gamma_{nm}+\alpha_{nm}\right]
\label{mtry}
\ee

where
\be
\alpha_{nm}=-\frac{2\rho}{z_n+z_m}
\sum_{p} \left[\Gamma_{ m p }\frac{X^{(p)}}{ X^{(m)}} -\Gamma_{ np }\frac{X^{(p)}}{ X^{(n)}}\right]\nonumber \\
\ee
and
\be
\gamma _{nm}=\frac{2\Gamma^{(nn)}+z_{n}-2\Gamma^{(mm)}-z_{m}}{z_{m}+z_{n}} 
\label{gdef}
\end{equation}

The second symmetry condition is Eq.(\ref{SII}) is
\be
{\bf{\tilde M}}^{A}= {\bf\Gamma}^{-1}\cdot {\bf{\tilde M}}\cdot {\bf z}-{\bf z}\cdot {\bf{\tilde M}^T}\cdot  [{\bf\Gamma}^T]^{-1}
\label{SIIa}
\ee

\section{The 2 Yukawa Case: Symmetric Matrix Results}
\label{symmetryiii}

We write  equation (\ref{aclos}) in matrix form \cite{blmhe2} 
\be
-{\overrightarrow{\bPi}_i}={\bf \tilde M}\cdot{\overrightarrow{\bf a}_i}
\label{M2eq}
\ee 
where
\be
{\overrightarrow{\bf X }_i}= \left[ 
\begin{array}{c}
X^{(1)}_i \\ 
X^{(2)}_i  \\
\end{array}
\right]\qquad
{\overrightarrow{\bPi}_i}=\left[ 
\begin{array}{c}
\Pi^{(1)}_i\\ 
\Pi^{(2)}_i \\
\end{array}
\right]\qquad
{\overrightarrow{\bf a}_i}=\left[ 
\begin{array}{c}
a^{(1)}_i \\ 
a^{(2)}_i \\
\end{array}
\right]
\label{xpiad}
\ee

Using the symmetry relation eq.(\ref{SI})
we get
\be
\left(\Gamma^{(12)}\frac{X^{(2)}}{X^{(1)}}-\Gamma^{(21)}\frac{X^{(1)}}{X^{(2)}}\right)= 
   \frac{s_{12}}{2}[ \chi_{12} -{{\gamma }_{12}} ]
\label{alfa1}
\ee
with
\be
\chi_{12}=\frac{{z_1} - {z_2}}
{z_1+z_2+2\,\Gamma^{(11)} + 2\,\Gamma^{(22)}}
\label{chi12}
\ee
\be
\gamma_{12}=\frac{z_1 - z_2 + 
    2\,{{\Gamma }^{(11)}} - 2\,{{\Gamma }^{(22)}}}{{z_1} + 
    {z_2}}
\ee
 in eq.(\ref{mtry}) we can write

\bear
{\bf \tilde M}=\frac{\rho}{2}\left[\begin{array}{cc}
X^{(1)}& 0\\
0 & X^{(2)}   \\
\end{array}
\right]
\left[\begin{array}{cc}
1 & 1-\chi_{12}\\ 
1+\chi_{12} & 1  \\
\end{array}
\right]
\left[\begin{array}{cc}
X^{(1)} & 0\\ 
0 & X^{(2)}   \\
\end{array}
\right]\nonumber \\
\label{mfin}
\eear
We rewrite eq.(\ref{M2eq}) as
\bear
2\left[\begin{array}{c}
{\cal G}^{(1)}\\
{\cal G}^{(2)}   \\
\end{array}
\right]
=\left[\begin{array}{cc}
1 & 1-\chi_{12}\\ 
1+\chi_{12} & 1  \\
\end{array}
\right]
\left[\begin{array}{c}
\beta^{(1)}\\
\beta^{(2)}   \\
\end{array}
\right]
\nonumber \\
\label{mfin2}
\eear
Here we have defined
\be
{\cal G}^{(1)}=\Gamma^{(11)}+\frac{X^{(2)}}{X^{(1)}} \Gamma^{(12)};\qquad
{\cal G}^{(2)}=\Gamma^{(22)}+\frac{X^{(1)}}{X^{(2)}} \Gamma^{(21)}
\label{calgd}
\ee
If we also define
\be
2{\cal G}^{(s)}={\cal G}^{(1)}+{\cal G}^{(2)};\qquad
2{\cal G}^{(12)}={\cal G}^{(1)}-{\cal G}^{(2)}
\label{calge}
\ee
then  we can solve eq.(\ref{mfin2})
\bear
 2{\cal G}^{(s)}=2\beta_s+\beta_{12}\chi_{12} ;\qquad
2{\cal G}^{(12)}=-\beta_s\chi_{12} ;
\eear
or
\bear
\beta_s=-2\frac{{\cal G}^{(12)}}{\chi_{12}} ;\qquad
\beta_{12}=\frac{2}{\chi_{12}}\left[ {\cal G}^{(s)}+ \frac{{\cal G}^{(12)}}{\chi_{12}}\right] ;
\label{betchi}
\eear

From the second symmetry condition Eq.(\ref{sIIa}) we get
\be
\frac{X^{(2)}}{X^{(1)}} z_1 \Gamma^{(12)}-\frac{X^{(1)}}{X^{(2)}} z_2 \Gamma^{(21)}-2 \Gamma^{(12)}\Gamma^{(21)}\chi_{12}+2 \tau_{12}=0
\label{sIIa}
\ee
where
\be
\tau_{12}=\left( 
    \frac{z_2\Gamma^{(11)}(z_1+\Gamma^{(11)}) -z_1\Gamma^{(22)}(z_2+\Gamma^{(22)})}
     {z_1+z_2+{2\,\Gamma^{(11)}} + {2\,\Gamma^{(22)}}} \right) 
 \label{tau12}
\ee
We remark also that
\begin{eqnarray}
\tau_{12}=\frac{1}{2}[z_2 \Gamma^{(11)}(1+\chi_{12})-z_1 \Gamma^{(22)}(1-\chi_{12})] \nonumber \\ 
+\chi_{12} \Gamma^{(11)}\Gamma^{(22)}
\label{tauy}
\end{eqnarray}
 in Eq.(\ref{sIIa}) we get
\bear
\frac{X^{(2)}}{X^{(1)}} z_1 \Gamma^{(12)}-\frac{X^{(1)}}{X^{(2)}} z_2 \Gamma^{(21)}+ 2\chi_{12} D_\Gamma\nonumber \\+[z_2 \Gamma^{(11)}(1+\chi_{12})-z_1 \Gamma^{(22)}(1-\chi_{12})]
=0
\label{sIIb}
\eear
Using now eq.(\ref{alfa1})\\
\bear
\frac{ X_{2}^2}{X_{1}^2}\Gamma_{12}^2-\frac{ X_{2}}{X_{1}}\Gamma_{12}\left[\frac{s_{12}\chi_{12}}{ 2}+z_2+2\Gamma^{(22)}\right]
+ \Gamma^{(22)}(z_2+\Gamma^{(22)})=0\nonumber \\
\label{gam12}
\eear

and
\bear
 \frac{ X_{1}^2}{X_{2}^2}\Gamma_{21}^2-\frac{ X_{1}}{X_{2}}\Gamma_{21}\left[-\frac{s_{12}\chi_{12}}{ 2}+z_1+2\Gamma^{(11)}\right]\nonumber \\
+ \Gamma^{(11)}(z_1+\Gamma^{(11)})=0\nonumber \\
\label{gam21}
\eear
from where
\bear
{\cal G}^{(1)}=\frac{1}{2}\left[\left\{\frac{s_{12}\chi_{12}}{ 2}+\frac{z_{12}}{\chi_{12}}\right\}-z_1-\sqrt{\Delta_\Gamma^{(2)}}\right]
\label{calg1}
\eear
\bear
{\cal G}^{(2)}=\frac{1}{2}\left[\left\{-\frac{s_{12}\chi_{12}}{ 2}+\frac{z_{12}}{\chi_{12}}\right\}-z_2-\sqrt{\Delta_\Gamma^{(1)}}\right]
\label{calg2}
\eear
with
\bear
\Delta_\Gamma^{(2)}=z_2^2+s_{12}\chi_{12}\left[\frac{s_{12}\chi_{12}}{ 4}+z_2+2\Gamma^{(22)}\right]
\eear
\bear
\Delta_\Gamma^{(1)}=z_1^2+s_{12}\chi_{12}\left[\frac{s_{12}\chi_{12}}{ 4}-z_1-2\Gamma^{(11)}\right]
\eear

\bear
\left[2 {\cal G}^{(1)}-\left\{\frac{s_{12}}{ 2}+z_{12}\right\}\chi_{12}+z_1\right]^2\nonumber \\-\left[2{\cal G}^{(2)}-\left\{-\frac{s_{12}}{ 2}+z_{12}\right\}\chi_{12}+z_2\right]^2=\Delta_\Gamma^{(2)}-\Delta_\Gamma^{(1)}=0
\nonumber \\
\label{chiq}
\eear

From here we get the equation
\bear
\left\{\chi_{12}-\frac{2 z_{12}}{s_{12}+2 {\cal G}_s}\right\}\left\{\chi_{12}-\frac{z_{12}+2 {\cal G}^{(12)}}{ s_{12}}\right\}=0 
\eear

which yields the two solutions 
\be
\chi_{12}=\frac{2 z_{12}}{s_{12}+2 {\cal G}_s}\quad (A);\qquad \chi_{12}=\frac{z_{12}+2 {\cal G}^{(12)}}{ s_{12}}\quad (B); 
\label{chi12s}
\ee
Notice first that in the zero density limit we get
\be
\chi_{12}\Longrightarrow\frac{2 z_{12}}{s_{12}};\qquad \chi_{12}\Longrightarrow\frac{z_{12}}{ s_{12}} 
\ee
and then in eqs(\ref{calg1}) and (\ref{calg2}) we get the correct zero density limit only from choice (B)
\bear
{\cal G}^{(1)}\simeq\frac{1}{2}\left[\left\{\frac{z_{12}}{ 2}+s_{12}\right\}-z_1-\frac{s_{12}}{2}\right]=0
\label{calg1s}
\eear
\bear
{\cal G}^{(2)}\simeq\frac{1}{2}\left[\left\{-\frac{z_{12}}{ 2}+s_{12}\right\}-z_2-\frac{s_{12}}{2}\right]=0
\label{calg2s}
\eear

Then
\bear 
\beta_s=-2{\cal G}^{(12)}\frac{s_{12}}{(z_{12}+2 {\cal G}^{(12)})};
\nonumber \\
\beta_{12}=2\frac{s_{12}}{(z_{12}+2 {\cal G}^{(12)})}
\left[ {\cal G}^{(s)}+ \frac{2{\cal G}^{(12)} s_{12}}{(z_{12}+2 {\cal G}^{(12)})}\right] ;\nonumber \\
\label{betaii}
\eear
These expressions turn out to be identical to those derived by Blum and Ubriaco using the diagonal approximation \cite{blmub}
\section{Thermodynamics by parameter integration }

We will use the notation and results of Blum and Hernando \cite{bluhe02}. We recall that

\be
 {\cal J}^{(n)}\Pi ^{(n)}=
{\cal I}^{(n)}
X^{(n)}
-\delta ^{(n)}
\label{s11fab} 
\ee

Remember that
\be
X^{(n)}=\gamma^{(n)}
+ \hat{\cal J}^{(n)} {\hat{B}}(z_n)
\label{eqfix} 
\ee
Here
\be
\hat{\cal J}^{(n)}=\delta 
\sigma\phi _{0}(z_{n}\sigma)
-2\rho \beta^0\sigma^3\psi _{1
}(z_{n})
\label{eq:s12faz} 
\ee
and 
\be
\hat{\gamma}^{(n)}=\delta^{(n)}-\frac{2 \beta^0}{z_n^2}\sum_\ell \rho 
\delta^{(n)}(1+\frac{z_n\sigma
}{2})
\label{eqs12fy} 
\ee

The total excess internal energy is
\begin{equation}
\frac{E(\beta)}{kTV}=\sum_{n}K_n\left\{\rho\delta^{(n)}{\hat{B}}
^{(n)}\right\} \label{equfi2}
\end{equation}
From eq.(\ref{pi2x})  we show that 
\begin{equation} 
-{\Pi}^{(n)}={\cal G}^{(n)}X^{(n)}
\label{tg2d}
\end{equation}
where \, ${\cal G}^{(n)}$ is a (generally algebraic) function of the coefficients $\bbeta\equiv \{\beta_1,\beta_2,..\}$. 
In fact   in eq.(\ref{s11fab}) 

\bear
\delta^{(n)}=\sum_{m}[\mathcal{M}^{nm}]X^{(m)}\nonumber \\
=\sum_{m}\{\mathcal{I}^{(n)}\delta_{nm}
+\mathcal{J}^{(n)}{\Gamma}^{(nm)}\}X^{(m)}\nonumber \\
=\mathcal{I}^{(n)}X^{(n)}
+\mathcal{J}^{(n)}\sum_{m}{\Gamma}^{(nm)}\}X^{(m)}\nonumber \\
=\{\mathcal{I}^{(n)}
+\mathcal{J}^{(n)}{\cal G}^{(n)}\}{X}^{(n)}
\eear
with

\be
{\cal G}^{(n)}=\sum_{m}\Gamma^{(nm)}\frac{X^{(m)}}{X^{(m)}}
\label{gcaldef}
\ee

For the 1 component case we get
\be
X^{(n)}=\frac{\delta^{(n)}}{{\cal I}^{(n)}+{\cal G}^{(n)}{\cal J}^{(n)}}
\label{calxx}
\ee

Then, since the 'charge' parameters are constants at constant temperature, the derivative of $\hat
{B}^{(n)}$ with respect to the scaling parameter ${\cal G}^{(n)}$   is
\bear
\left[\frac{\partial\hat{B}^{(n)}}{\partial{\cal G}^{(n)}}\right]= \left[  \mathcal{J}^{(n)}\right]  ^{-1}\left\{
\frac{\partial\left( {X}^{(n)}\right)  }{\partial {\cal G}^{(n)}}\right\}\nonumber \\ = -\left[  \mathcal{J}^{(n)}\right]  ^{-1}\left[\frac{\delta^{(n)} \mathcal{J}^{(n)}}{({\cal I}^{(n)}+{\cal G}^{(n)}}\right]
\eear
where we use the fact that $ \mathcal{J}^{(n)}$ is independent of ${\cal G}^{(n)}$. \\
The desired energy derivative Eq.(\ref{equfi2})are
\begin{equation}
\frac{\partial E}{\partial{\cal G}^{(n)}}= - \rho[{X}^{(n)}]^2
\ee
or

\bear
\frac{\partial E}{\partial{\cal G}^{(s)}}= - \sum_n\rho[{X}^{(n)}]^2
;\qquad
\nonumber \\ \frac{\partial E}{\partial{\cal G}^{(nm)}}= - \rho\{[{X}^{(n)}]^2-[{X}^{(m)}]^2\}
\label{meqz2p1}
\eear
The  integrability condition is satisfied since
\bear
\frac{\partial^2 E}{\partial{\cal G}^{(n)}\partial{\cal G}^{(m)} }=\frac{\partial^2 E}{\partial{\cal G}^{(m)}\partial{\cal G}^{(n)} }=\nonumber \\
=\delta^{Kr}_{nm} \left[ 2 \rho[{X}^{(n)}]^2\frac{\mathcal{J}^{(n)}}{{\cal I}^{(n)}+{\cal G}^{(n)}{\cal J}^{(n)}}\right]\nonumber \\
\label{crossder}
\eear

We use now Eq.(\ref{beta1c}) to obtain
\bear
\frac{\partial E}{\partial{\cal G}^{(s)}}=\frac{1}{2}\left[\beta _s^2+s_{12}
\beta _s+z_{12}\beta _{12} 
\right]=
\eear
\bear
\qquad=\frac{s_{12} z_{12}}{2(2{\cal G}^{(12)}+z_{12})}\left[2{\cal G}^{(s)}+s_{12}-\frac{s_{12} z_{12}}{(2{\cal G}^{(12)}+z_{12})}\right]
\label{calgsd}
\eear
and
%\begin{widetext}
\bear
\frac{\partial E}{\partial{\cal G}^{(12)}}=\frac{1}{2}\left[\beta_s (\beta_s+s_{12})+z_{12}
\beta _s+\frac{z_{12}}{2 s_{12}}\{\beta_s^2-\beta _{12}^2\} 
\right]=\nonumber \\
=\frac{s_{12} z_{12}}{4(2{\cal G}^{(12)}+z_{12})^2}\left[s_{12}^2+z_{12}^2-\left\{2{\cal G}^{(s)}+s_{12}-\frac{2s_{12} z_{12}}{(2{\cal G}^{(12)}+z_{12})}\right\}^2\right]\nonumber \\
-\frac{s_{12}z_{12}}{4};\nonumber \\
\label{calg12d}
\eear
%\end{widetext}
Thermodynamic integration of these equations leads to
\begin{widetext}

\bear
\Delta S=-\frac{k}{2\pi}?? \nonumber \\\left[\left(\frac{1}{8}\frac{s_{12}z_{12}}{(z_{12}+2{\cal G}^{(12)})}\right)^3\left\{\frac{ 1}{3}+\left(1-\frac{(z_{12}+2{\cal G}^{(12)})(s_{12}+2{\cal G}^{(s)})}{s_{12}z_{12}}\right)^2\right\}-\left(\frac{s_{12}z_{12}}{8}\right) \left(\frac{s_{12}^2+z_{12}^2}{(z_{12}+2{\cal G}^{(12)})}-z_{12}+2{\cal G}^{(12)}\right) 
+\frac{s_{12}^3}{12}\right]\nonumber \\
\label{s3symg}
\eear
\end{widetext}

\be
\Delta S=-\frac{k}{2\pi} \left[\frac{ \beta_s^3}{6}+\frac{ \beta_s}{4}[(\beta_s s_{12}+\beta_{12}z_{12}]-
\frac{z_{12}^2 (\beta_s^2-\beta_{12}^2)}{8(\beta_s+s_{12})}\right]
\label{s3finsym}
\ee
Although the derivation of this equation is completely different from that obtained using the diagonal $\bGamma$ assumption, the resulting entropy is identical to this case when proper reference states are used \cite{lin02}.\\

\bibliographystyle{plain}
\bibliography{}

\end{document}